\documentclass{Interspeech}
\usepackage{graphicx}
\usepackage{amsmath}
\usepackage{multirow}
\usepackage[table,xcdraw]{xcolor}
\usepackage{graphicx}
\usepackage{multicol} % For multiple columns
\usepackage[noadjust]{cite}
\usepackage{amsmath}
\usepackage{ulem}
\usepackage{amssymb}
\usepackage{hyperref}
\usepackage{color}
\usepackage{booktabs} % For \cmidrule
\usepackage{multirow}
\usepackage{float}
\usepackage{verbatim}
\usepackage{mathtools}
\usepackage{pifont}
\usepackage{booktabs}
\usepackage{algorithm}
\usepackage{algorithmic}
\newcommand{\cmark}{\ding{51}}  % Checkmark
\newcommand{\xmark}{\ding{55}}  % X mark

\interspeechcameraready

\title{ATMM-SAGA: Alternating Training for Multi-Module with Score-Aware Gated Attention SASV system}

\author[affiliation={1}]{Amro}{Asali}
\author[affiliation={1}]{Yehuda}{Ben-Shimol}
\author[affiliation={2,3}]{Itshak}{Lapidot}
\affiliation{Electrical and computer engineering school}{Ben Gurion University of the Negev}{Israel}
\affiliation{Electrical engineering school}{Afeka the Academic College of Engineering in Tel Aviv}{Israel}
\affiliation{}{Avignon University}{LIA, France}
\email{asaliam@bgu.ac.il, benshimo@bgu.ac.il, itshakl@afeka.ac.il}


\begin{document}
\maketitle
\begin{abstract}
%996 charachters
The objective of \textit{automatic speaker verification} (ASV) systems is to determine whether a given test speech utterance corresponds to a claimed enrolled speaker. These systems have a wide range of applications, and ensuring their reliability is crucial. In this paper, we propose a \textit{spoofing-robust automatic speaker verification} (SASV) system employing a \textit{score-aware gated attention} (SAGA) fusion scheme, integrating scores from a pre-trained countermeasure (CM) with speaker embeddings from a pre-trained ASV. Specifically, we employ the AASIST and ECAPA-TDNN models. SAGA acts as an adaptive gating mechanism, where the CM score determines how strongly ASV embeddings influence the final SASV decision. Experiments on the ASVspoof2019 \textit{logical access} dataset demonstrate that the proposed SASV system achieves an SASV \textit{equal error rate} (SASV-EER) and \textit{agnostic detection cost function} (a-DCF) of 2.31\%, 0.0603 for the development set and 2.18\%, 0.0480 for the evaluation set.
\end{abstract}

\section{Introduction}
Recent studies have demonstrated that ASV systems are undergoing a gradual evolution, acquiring the capacity to reject spoofed inputs in a zero-shot manner. However, rapid advancements in speech synthesis techniques, such as \textit{text-to-speech} (TTS) or \textit{voice conversion} (VC), highlight the ongoing necessity to further enhance spoofing-robust ASV systems ~\cite{jung2024extentasvsystemsnaturally}. The findings indicate that contemporary SASV systems exhibit superior performance in terms of the SASV-EER, as measured by trials encompassing all three classes: target, bona fide non-target, and spoofed non-target \cite{jung2022sasv2022spoofingawarespeaker}. 

Reliable speaker verification is frequently comprised of two distinct subsystems: ASV~\cite{ggemini,Desplanques-2020} and spoofing CM~\cite{jung2021aasistaudioantispoofingusing,tak2021end,borodin24_asvspoof} classifiers. This prompts the following research question: how should the two subsystems be integrated to achieve robust speaker verification? The results presented in~\cite{ge2022potentialjointlyoptimised} indicate that while joint optimization enhances the reliability of ASV at the SASV level, superior performance is achieved by integrating fixed pre-trained subsystems. The CM and ASV can be combined in a cascade or in a parallel fashion~\cite{sahidullah2016integrated,avishai,weizman2024tandemspoofingrobustautomaticspeaker}, with integration typically occurring at the score or embedding levels, as evidenced in~\cite{probfusi,choi22b-interspeech,Liu-2024}.

\section{Background}
This section presents a concise review of the pertinent literature on the proposed solution, along with a concise overview of the ASV and CM systems employed for embedding extraction.
\subsection{Automatic speaker verification system}
In this study, the \textit{emphasized channel attention, propagation, and aggregation time delay neural network} (ECAPA-TDNN) speaker verification system is employed~\cite{Desplanques-2020}. The system utilizes 80-dimensional \textit{Mel-frequency cepstral coefficients} (MFCCs) as features and incorporates a modified Res2Net as its backbone processing block, augmented with dimensional \textit{squeeze-excitation} (SE) blocks, to model global channel interdependencies. The model incorporates attentive statistics pooling, enabling the model to select relevant frames, and multi-layer feature aggregation, which captures both shallow and complex speaker identity features at the frame level and aggregates them into utterance-level embeddings.

\subsection{CM system}
In this study, we adopt the \textit{audio anti-spoofing using integrated spectro-temporal graph attention networks} (AASIST) approach, which utilizes the RawNet2 frontend and graph attention framework \cite{jung2021aasistaudioantispoofingusing}. The model employs a Sinc convolution encoder to extract time-frequency representations, which are then processed by a residual network to learn high-level features. Subsequently, two graph modules are employed to model the spectral and temporal domains, respectively.

\subsection{Structural transformation on ReLU}
For the affine layer $W_i x + b_i$, with learnable weights and biases $W_i$ and $b_i$, respectively, we define a structural transformation $t\mathrm{ReLU}$ as its activation function. $t\mathrm{ReLU}$ is defined as follows:
\begin{equation}  
t\mathrm{ReLU}_{W_a}(W_i x + b_i) =
\max\left(W_a\left(W_i x + b_i\right), \mathbf{0}\right)
\label{equation:frelu}
\end{equation}
where $W_a$ is the learnable structural transformation, initialized as the identity matrix. Here, the $\max(\cdot,\cdot)$ operation is performed element-wise. Same definition with a diagonal constraint of $W_a$ was implemented in~\cite{pact,Bell-2021,Liu-2024}. 

\section{Proposed system}
In this section, we will present the operation of the proposed system, which has been designed to address the SASV problem. Given a pair of utterances, an enrollment utterance $U_{\mathrm{erl}}$ of the target speaker and a test utterance $U_{\mathrm{tst}}$, the system will evaluate whether $U_{\mathrm{tst}}$ was spoken by the target speaker (output $y=1)$ or by a non-target speaker ($y=0$). Non-target attacks can be either zero-effort impostor attacks or spoofing attacks.

\vfill
\noindent\hrulefill

\href{https://github.com/amro-asali-2/ATMM-SAGA}{https://github.com/amro-asali-2/ATMM-SAGA}

\subsection{Model architecture}
\begin{figure}[t]
   \centering
   \includegraphics[width=\linewidth]{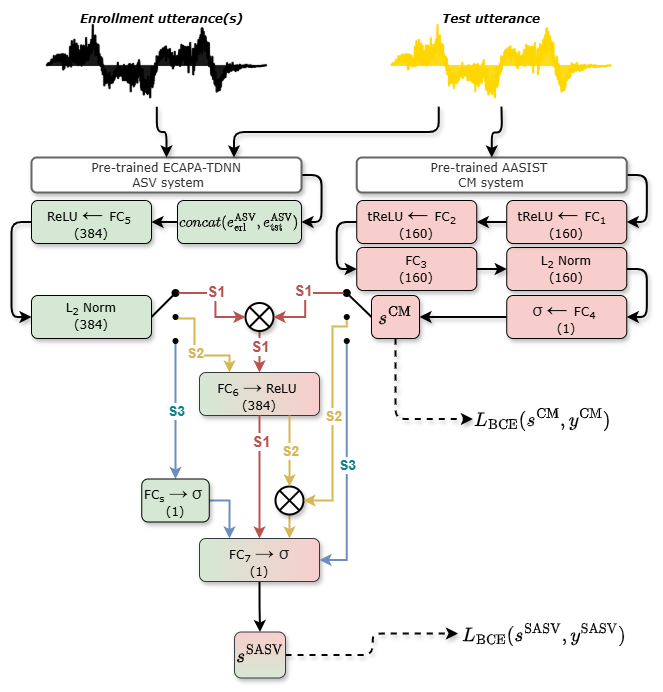}
   \caption{Diagram of the proposed SASV system, illustrating various CM score integration strategies, represented by distinct color-coded paths. Dashed arrows denote operations exclusive to the training stage.}
   \label{fig:propsedsystem}

\end{figure}

This study investigates the integration of speaker embeddings, derived from a pre-trained ECAPA-TDNN model \cite{Desplanques-2020}, with a CM score, based on a pre-trained AASIST embeddings \cite{jung2021aasistaudioantispoofingusing}. This integration is achieved by multiplying the obtained CM score with the normalized activations of the speaker embeddings, thereby generating spoofing-aware speaker embeddings. This approach is referred to as \textit{score-aware gated attention} (SAGA). These embeddings are subsequently employed to calculate SASV scores. The proposed system is illustrated in~\autoref{fig:propsedsystem}. In this investigation two integration levels are examined while maintaining the integrity of the overall system architecture. Additionally, a score fusion strategy is explored to facilitate a more comprehensive comparative analysis.

\subsubsection{Strategy S1: Early Integration}
As depicted in~\autoref{fig:propsedsystem}, S1 follows the red path, where the AASIST's CM embeddings are processed through two fully connected layers on the top right processing path (colored soft red in the figure). These embeddings undergo parameter-shared $t\mathrm{ReLU}$ activations, as defined in \autoref{equation:frelu}. Subsequently, the embeddings pass through an additional fully connected layer and are normalized using L2-normalization, and are then processed through a final fully connected layer with a Sigmoid activation, ultimately yielding a CM score, $\mathrm{s^{CM} \in [0,1]}$.

Concurrently, on the left processing path (colored green in the figure), the ECAPA-TDNN speaker embeddings are concatenated and processed through a fully connected layer, followed by the classic ReLU activation function. These embeddings are then normalized using L2-normalization to obtain the normalized speaker embeddings, ${\mathbf{e}^{\mathrm{ASV}}}$.
To incorporate SAGA, a multiplicative gating mechanism is applied, in which the CM score functions as an adaptive attention weight over the ASV embeddings:
\begin{equation}
\mathbf{e}^\mathrm{SASV} = g(s^\mathrm{CM}, \mathbf{e}^\mathrm{ASV}) = s^\mathrm{CM}\mathbf{e}^\mathrm{ASV}
\end{equation} where
 ${g(s^\mathrm{CM}, \mathbf{e}^\mathrm{ASV})}$ represents the SAGA operation, ensuring that spoofed samples (${s^\mathrm{CM}\approx 0 }$) are suppressed while bona fide samples (${s^\mathrm{CM}\approx1 }$) are preserved. $\mathbf{e}^\mathrm{SASV}$ represents the spoofing-aware speaker verification embeddings.
These spoofing-aware embeddings undergo further processing through an additional fully connected layer with ReLU activation before being processed through a final fully connected layer with a Sigmoid activation, ultimately yielding an SASV score.

\subsubsection{Strategy S2: Late Integration}
While S1 and S2 strategies share the same overall architecture, the approach of S2 involves delaying the application of the SAGA mechanism, integrating it later in the ASV processing path (yellow path in~\autoref{fig:propsedsystem}).
\subsubsection{Strategy S3: Score Fusion}  
In this strategy, score fusion between the ASV and CM systems is achieved through a fully connected layer to generate the SASV score. This approach (blue path in~\autoref{fig:propsedsystem}) is used exclusively for comparative analysis.
\subsection{Training}
 In order to train the model for both speaker verification and countermeasure tasks, a multi-task learning paradigm is adopted as outlined in \cite{Li2020JointDecision}. During the training phase of our proposed system, the total multi-task classification loss is defined as follows:  
\begin{equation}
    L^{\mathrm{total}} = \lambda \cdot L^{\mathrm{SASV}}_{\mathrm{BCE}}(s^\mathrm{SASV},y^\mathrm{SASV}) + (1 - \lambda) \cdot L^{\mathrm{CM}}_{\mathrm{BCE}}(s^\mathrm{CM},y^\mathrm{CM})
    \label{equation:total_loss}
\end{equation}
where $\lambda \in [0, 1]$ is fixed at 0.5 to assign equal importance to both tasks. In this context, $L^{\mathrm{SASV}}_{\mathrm{BCE}}(\cdot,\cdot)$ denotes the \textit{binary cross-entropy} (BCE) loss calculated between the system's SASV output score $s^\mathrm{SASV}\in[0,1]$ and the SASV label $y^\mathrm{SASV}\in\{0,1\}$. In contrast, $ L^{\mathrm{CM}}_{\mathrm{BCE}}(\cdot,\cdot)$ represents the BCE loss calculated between the system's CM output score $\mathrm{s^{CM}}\in[0,1]$ and the CM label ${y^\mathrm{CM}\in\{0,1\}}$, where ${y^\mathrm{CM}=1}$ indicates a bona fide trial and $y^\mathrm{CM}=0$ represents a spoof trial.

%Where $y^{SASV}=1$ corresponds a target trial, while $y^{SASV}=0$ is associated with either a non-target trial, or a spoof trial
\subsubsection{Alternating Training for Multi-Module (ATMM)}
During the training phase, two separate datasets are employed, obtained by partitioning the unified training set used for training the SAGA SASV system without employing ATMM, as described in~\autoref{table:training data}. The first dataset is only used for the spoofing countermeasure training, while the second dataset is only used for the speaker verification training. For the CM training dataset, the ASVspoof2019 \textit{logical access} (LA) train set \cite{wang2020asvspoof2019largescalepublic} is employed to generate pairs of enrollment and test utterances. Specifically, each bona fide utterance is paired with a random subset of the same speaker's bona fide utterances to form target trials and with other speakers' bona fide utterances to form zero-effort non-target trials. Additionally, we have paired bona fide utterances with the same speaker's spoofed utterances to create spoof non-target trials.
\begin{table}[t]
\caption{Overview of the training datasets used in our experiments, detailing the number of target, non-target, and spoofed trials for both the spoofing CM and speaker verification datasets.}
\label{table:training data}
  \centering
  \begin{tabular}{lccc}
    \toprule
    \textbf{Dataset}   & \textbf{Target} & \textbf{Non-target} & \textbf{Spoof} \\ \midrule
    \\
    {Spoofing CM} & 262228    & 249094 & 463910
    \\
    {Speaker verification}  & 806025    & 779601 & 0
    \\
    \bottomrule
  \end{tabular}
  \label{traindstable}

\end{table}

The speaker verification training dataset is comprised of the VoxCeleb1 E and H partitions, and the ASVspoof2019 LA training set bona fide pairs. The ATMM algorithm optimizes joint training by alternating update focus between the two modalities. At each training step, a binary decision determines whether to prioritize the ASV or CM module. This prevents overfitting to either task, while preserving previously learned knowledge with selective backpropagation achieved through strategic weight freezing. It is important to note that a pair of utterances (enrollment and test) is required for both CM and ASV training, since the joint representation layers (depicted with a mix of soft red and green in~\autoref{fig:propsedsystem}) remain unfrozen, ensuring continuous learning. Maintaining $\lambda \in (0,1)$, empirically chosen to alternate between 0.1 and 0.9, allows for continuous adaptive gradient flow from both tasks, preserving joint feature learning while preventing overfitting. A detailed description of the training algorithm is provided in Alg.\autoref{alg:training_algorithm}.

\begin{algorithm}[t]
\caption{ATMM: One Round of Training}
\label{alg:training_algorithm}
\begin{algorithmic}[1]
\FOR{$100$ iterations}
    \STATE Choose a random $p \in \{0, 1\}$.
    \IF{$p = 0$}
        \STATE Set $\lambda \gets 0.1$.
        \STATE Sample $1\%$ of the spoofing CM dataset.
        \STATE Freeze the speaker verification weights (colored in green), as shown in Figure~\ref{fig:propsedsystem}.  
    \ELSE
        \STATE Set $\lambda \gets 0.9$.
        \STATE Sample $1\%$ of the speaker verification dataset.
        \STATE Freeze the CM weights (colored in soft red), as shown in Figure~\ref{fig:propsedsystem}.
    \ENDIF
    \STATE Compute the total loss according to \autoref{equation:total_loss}.
    \STATE Perform backpropagation and update the weights of the unfrozen components.
\ENDFOR
\end{algorithmic}
\end{algorithm}

\section{Experimental setup}
In the following, we describe the experimental setup for our proposed SASV system, detailing the datasets used for evaluation and the metrics employed to assess performance.

\subsection{Dataset}
The ASVspoof 2019 LA dataset is a widely used benchmark~\cite{wang2020asvspoof2019largescalepublic}. This dataset comprises genuine speech utterances and those that have been spoofed using a variety of text-to-speech (TTS) and voice conversion (VC) techniques. The dataset is split into three sets: training (Train), development (Dev), and evaluation (Eval). It includes official development and evaluation protocols, and for each trial, multiple corresponding enrollment utterances are provided to register the target speaker. Notably, the training and development sets incorporate the same six spoofing attacks (A01–A06), which include four TTS and two VC attacks. In contrast, the evaluation set comprises 11 previously unseen attacks (A07–A15, A17, A18) and two additional attacks (A16, A19) that, despite employing similar underlying algorithms as some training attacks, are trained with different data. The evaluation set includes 5,370 target trials, 33,327 non-target trials, and 63,882 spoofed trials.

\subsection{Evaluation Metrics}
\subsubsection{SASV-EER}
The SASV-EER evaluates system performance across all three trial types: target, zero-effort non-target, and spoofed non-target. It is defined as the error rate at the threshold where the false rejection rate of target trials equals the false acceptance rate of both zero-effort and spoofed non-target trials, providing a unified measure of robustness against both speaker mismatch and spoofing attacks.

\subsubsection{Minimum Normalized Agnostic Detection Cost Function (min a-DCF)}

The min a-DCF evaluates system performance by optimizing the Bayes risk while incorporating class priors and detection costs \cite{adcf}:

\begin{equation}
\min_t \frac{ C_{\text{miss}}\pi_{\text{tar}}P_{\text{miss}}(t) + C_{\text{fa,non}}\pi_{\text{non}}P_{\text{fa,non}}(t) + C_{\text{fa,spf}}\pi_{\text{spf}}P_{\text{fa,spf}}(t)}{\min \left\{ C_{\text{miss}} \pi_{\text{tar}}, C_{\text{fa,non}} \pi_{\text{non}} + C_{\text{fa,spf}} \pi_{\text{spf}} \right\}
}
\end{equation} where \( P_{\text{miss}}, P_{\text{fa,non}}, P_{\text{fa,spf}} \) are the miss rate (false rejection), non-target false acceptance rate, and spoof false acceptance rate at threshold \( t \), respectively. The terms \( \pi_{\text{tar}}, \pi_{\text{non}}, \pi_{\text{spf}} \) denote the priors for target, non-target, and spoof trials, while \( C_{\text{miss}}, C_{\text{fa,non}}, C_{\text{fa,spf}} \) are their respective detection costs.

Unlike the EER metric, which operates independently of class priors and decision costs, min a-DCF provides a more comprehensive evaluation by considering the trade-off between different types of errors under real-world conditions.

\section{Results}
In this section, we present a summary of our experimental findings. 
The present study commences with an examination of the influence of distinct training methodologies on model performance and its generalizability to unseen attacks. 
In the subsequent step, an evaluation of the various strategies for integrating the CM score is conducted, with a focus on the effectiveness of the proposed SASV system in comparison with several baseline methods. 
The assessment of performance is conducted through the utilization of the SASV-EER and min a-DCF methodologies. 
Confidence intervals are computed via bootstrapping, utilizing 1,000 iterations and a 95\% confidence level~\cite{Confidence_Intervals}.
\subsection{Comparison of Training Approaches}

In order to enhance the robustness of the SASV system, a series of experiments were conducted, in which various training approaches were implemented in conjunction with the S1 integration strategy. These experiments involved modifying the training algorithm, as detailed in Alg. \autoref{alg:training_algorithm}. In addition, we incorporated regularization techniques including \textit{dropout} and \textit{batch normalization} (BN) layers to mitigate the risk of overfitting and stabilize the training process. By systematically evaluating different training configurations, we aimed to identify the most effective combination for enhancing model generalization.

\begin{table}[t]
\caption{\small Performance comparison of different training technique combinations with S1 integration strategy, evaluated in terms of min a-DCF and SASV-EER, with confidence intervals included, on the ASVspoof2019 LA dataset’s development and evaluation sets.}
\label{table:train_tech}
    \centering
    \scriptsize
    \begin{tabular}{c c c c c c c}
        \toprule
        \multicolumn{2}{c}{\textbf{min a-DCF}} & \multicolumn{2}{c}{\textbf{EER (\%)}} & \textbf{BN} & \textbf{Drop} & \textbf{ATMM} \\
        \textbf{Dev} & \textbf{Eval} & \textbf{Dev} & \textbf{Eval} & & &  \\ 
        \midrule
        $\mathbf{0.0500}$ & 0.1464 & $\mathbf{1.46}$ & 5.74 & \textcolor{orange}{\xmark} & \textcolor{orange}{\xmark} & \textcolor{orange}{\xmark} \\
        \scalebox{0.6}{[0.0409, 0.0567]} & \scalebox{0.6}{[0.1413, 0.1517]} &\scalebox{0.6}{[1.27, 1.77]} &\scalebox{0.6}{[5.61, 5.90]} & & & \\
        0.1189 & 0.1386 & 4.45 & 5.22 & \cmark & \xmark & \xmark \\
        \scalebox{0.6}{[0.1174, 0.1430]} & \scalebox{0.6}{[0.1321, 0.1439]} & \scalebox{0.6}{[3.39, 5.17]} &\scalebox{0.6}{[5.01, 5.47]} & & & \\
        0.0653 & 0.1422 & 2.25 & 5.58 & \xmark & \cmark & \xmark \\
        \scalebox{0.6}{[0.0552, 0.0730]} & \scalebox{0.6}{[0.1364, 0.1466]} & \scalebox{0.6}{[1.83, 2.55]}&\scalebox{0.6}{[5.38, 5.69]} & & & \\
        0.1079 & 0.1315 & 4.48 & 4.98 & \cmark & \cmark & \xmark \\
        \scalebox{0.6}{[0.0933, 0.1185]} & \scalebox{0.6}{[0.1243, 0.1363]} & \scalebox{0.6}{[3.41, 5.16]} &\scalebox{0.6}{[4.77, 5.15]} & & & \\
        0.0603 & $\mathbf{0.0480}$ & 2.31 & $\mathbf{2.18}$ & \textcolor{blue}{\xmark} & \textcolor{blue}{\xmark} & \textcolor{blue}{\cmark} \\
        \scalebox{0.6}{[0.0522, 0.0609]} & \scalebox{0.6}{[0.0435, 0.0527]} &\scalebox{0.6}{[1.63, 3.15]} &\scalebox{0.6}{[1.81, 2.56]} & & & \\
        0.0975 & 0.0702 & 6.60 & 4.21 & \cmark & \xmark & \cmark \\
        \scalebox{0.6}{[0.0863, 0.1094]} & \scalebox{0.6}{[0.0644, 0.0754]} &\scalebox{0.6}{[5.48, 7.59]} &\scalebox{0.6}{[3.77, 4.57]} & & & \\
        0.0620 & 0.0516 & 2.42 & 2.27 & \xmark & \cmark & \cmark \\
        \scalebox{0.6}{[0.0519, 0.0699]} & \scalebox{0.6}{[0.0474, 0.0554]} & \scalebox{0.6}{[1.66, 3.31]} &\scalebox{0.6}{[1.98, 2.60]} & & & \\
        0.0937 & 0.0707 & 6.84 & 4.17 & \cmark & \cmark & \cmark \\
        \scalebox{0.6}{[0.0818, 0.1051]} & \scalebox{0.6}{[0.0648, 0.0752]} &\scalebox{0.6}{[5.74, 7.87]} &\scalebox{0.6}{[3.79, 4.49]} & & & \\
        \bottomrule
    \end{tabular}
\end{table}

\begin{figure}[t]
  \centering
  \includegraphics[width=\linewidth]{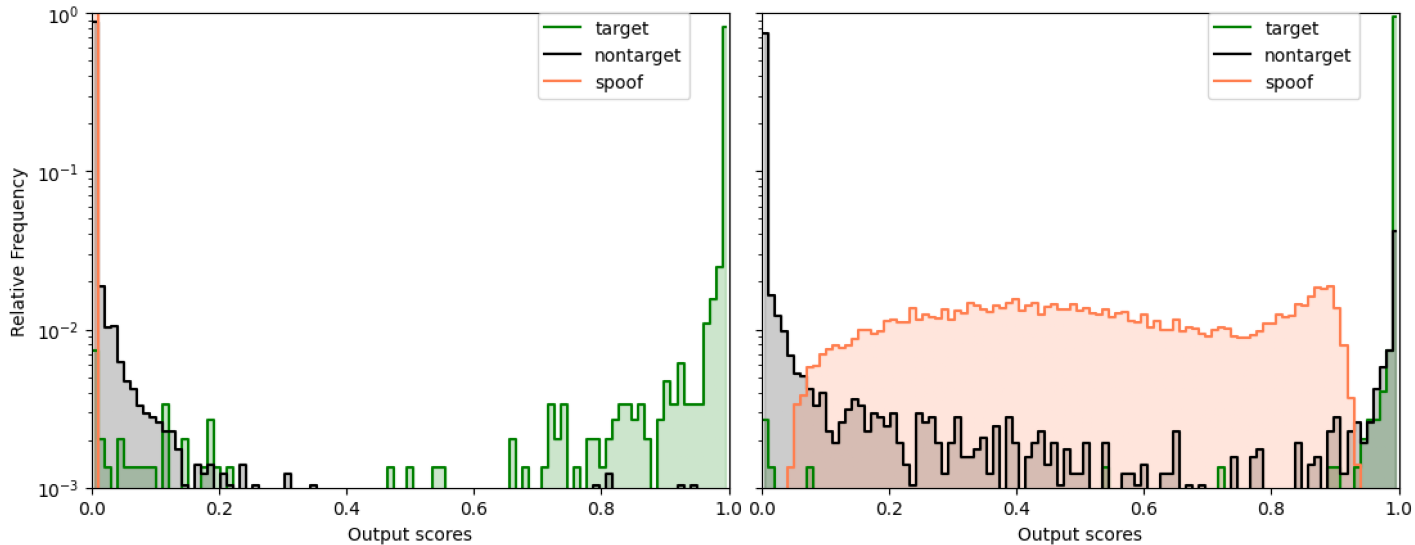}
    \caption{Log-scale normalized histograms of output scores for target, zero-effort non-target, and spoof trials on the ASVspoof2019 LA development dataset. The left plot represents results from a system trained using conventional training (without ATMM), while the right plot corresponds to the system trained with ATMM. Both systems utilize the S1 strategy.}

  \label{fig:dev_scores}
\end{figure}

The results in \autoref{table:train_tech} indicate that while conventional regularization techniques, such as BN and dropout, have been effective in some cases, modifying the training procedure, as detailed in Alg.\autoref{alg:training_algorithm}, leads to significant improvements. Notably, applying ATMM without BN and without \textit{dropout} (colored in blue in~\autoref{table:train_tech}) resulted in the lowest min a-DCF and SASV-EER scores on the evaluation set. However, it performed worse on the development set compared to the model without ATMM (colored in orange in~\autoref{table:train_tech}). This discrepancy can be attributed to the development set containing the same attacks as the training set. This suggests that a model trained with ATMM is less prone to overfitting to training attacks and generalizes better to unseen attacks. Figure~\ref{fig:dev_scores} further suggests that ATMM prevents overfitting by balancing score distributions. Without ATMM (left side), spoof scores are tightly clustered near zero, indicating overconfidence. After employing ATMM (right side), spoof scores are more evenly distributed while maintaining a clear separation from target scores. This suggests that careful adjustment of the training algorithm can be more effective in reducing overfitting and enhancing model performance than standard regularization methods.

\subsection{Comparison of CM Score Integration Strategies and Baseline Systems}
In this subsection, we analyze the optimal strategy for integrating the CM score into the SASV system and compare the proposed solution with fusion-based baseline systems from the SASV2022 challenge~\cite{jung2022sasv2022spoofingawarespeaker} and the multi-task-based G-SASV system from~\cite{Liu-2024}. As shown in \autoref{tab:experiment_results}, SAGA outperforms score fusion, with the S1 strategy achieving slightly better performance than S2, indicating that early integration is a better strategy than late integration. To further demonstrate the efficacy of the proposed approach, we compare it against individual ASV and CM systems, which are utilized for feature extraction. While the ASV and CM subsystems achieve state-of-the-art performance on their respective tasks, they prove ineffective when applied individually to the spoofing-robust automatic speaker verification task, as reported in~\cite{Liu-2024,probfusi}. As demonstrated in \autoref{tab:experiment_results}, the confidence intervals for the employed evaluation metrics on the evaluation dataset for the proposed solution lie entirely below those of the baseline and individual systems. This indicates that integration using the ATMM-SAGA training and integration framework yields statistically significant improvements over both embeddings and score fusion.

\begin{table}
\caption{Comparison of different CM score integration strategies with SASV baselines, standalone ASV and CM systems, evaluated in terms of min a-DCF and SASV-EER, with confidence intervals included, on the ASVspoof2019 LA dataset’s development and evaluation sets.}
\label{tab:experiment_results}
\centering
\footnotesize
\begin{tabular}{lcccccc}
\toprule
\textbf{Systems} & \multicolumn{2}{c}{\textbf{SASV-EER (\%)}} & \multicolumn{2}{c}{\textbf{min a-DCF}} \\
& Dev  & Eval & Dev & Eval \\
\midrule
ECAPA-TDNN \cite{Desplanques-2020}  & 17.31  & 23.84 &- &- \\
AASIST \cite{jung2021aasistaudioantispoofingusing}       & 15.86  & 24.38 &- &- \\
Baseline1 \cite{jung2022sasv2022spoofingawarespeaker}    & 13.06  & 19.31 &- &- \\
G-SASV \cite{Liu-2024} &- & 8.62 &- &- \\
Baseline2 \cite{jung2022sasv2022spoofingawarespeaker}    & 3.10   & 6.54 &- &-\\
S3     & 3.87 & 5.45 & 0.1087 & 0.1245 \\
    &\scalebox{0.6}{[3.40, 4.79]} &\scalebox{0.6}{[4.95, 5.87]} & \scalebox{0.6}{[0.0942, 0.1198]} & \scalebox{0.6}{[0.1127, 0.1354]}\\
\midrule
S1     & 2.31 & $\mathbf{2.18}$ & 0.0603 & $\mathbf{0.0480}$\\
     &\scalebox{0.6}{[1.63, 3.15]} &\scalebox{0.6}{[1.81, 2.56]} & \scalebox{0.6}{[0.0522, 0.0609]} & \scalebox{0.6}{[0.0435, 0.0527]}\\
S2     & $\mathbf{2.28}$ & 2.19 & $\mathbf{0.0571}$ & 0.0501   \\
     &\scalebox{0.6}{[1.62, 3.17]} &\scalebox{0.6}{[1.86, 2.48]} & \scalebox{0.6}{[0.0490, 0.0651]} & \scalebox{0.6}{[0.0454, 0.0535]}\\
\bottomrule
\end{tabular}
\end{table}

\section{Conclusions and Future Work}

This paper presents a robust SASV system that integrates CM scores with speaker embeddings using the SAGA mechanism and ATMM algorithm. The proposed approach enables seamless fusion of ASV and CM pre-trained models while maintaining a compact and efficient structure. Our results demonstrate that SAGA is the superior method for incorporating CM scores, significantly outperforming conventional score fusion. Furthermore, applying the SAGA mechanism early in the network enhances the system’s ability to discriminate between bona fide and spoofed samples. We also show that ATMM surpasses conventional regularization techniques, such as BN and \textit{dropout}, in mitigating overfitting and improving generalization. By alternating between ASV and CM training, ATMM effectively balances modality learning and strengthens feature discrimination, leading to superior SASV performance. Despite these advancements, there remains considerable room for further optimization. Future work may focus on refining the system by enhancing the SAGA mechanism, optimizing the ATMM algorithm, and exploring complementary feature spaces.

\ifinterspeechfinal
\section{Acknowledgments}
This work is supported by the Israel Innovation Authority under project numbers 82457 and 82458.
\fi
\bibliographystyle{IEEEtran}
\bibliography{references}

\begin{thebibliography}{10}
\providecommand{\url}[1]{#1}
\csname url@samestyle\endcsname
\providecommand{\newblock}{\relax}
\providecommand{\bibinfo}[2]{#2}
\providecommand{\BIBentrySTDinterwordspacing}{\spaceskip=0pt\relax}
\providecommand{\BIBentryALTinterwordstretchfactor}{4}
\providecommand{\BIBentryALTinterwordspacing}{\spaceskip=\fontdimen2\font plus
\BIBentryALTinterwordstretchfactor\fontdimen3\font minus \fontdimen4\font\relax}
\providecommand{\BIBforeignlanguage}[2]{{%
\expandafter\ifx\csname l@#1\endcsname\relax
\typeout{** WARNING: IEEEtran.bst: No hyphenation pattern has been}%
\typeout{** loaded for the language `#1'. Using the pattern for}%
\typeout{** the default language instead.}%
\else
\language=\csname l@#1\endcsname
\fi
#2}}
\providecommand{\BIBdecl}{\relax}
\BIBdecl

\bibitem{jung2024extentasvsystemsnaturally}
J.~weon Jung, X.~Wang, N.~Evans, S.~Watanabe, H.~jin Shim, H.~Tak, S.~Arora, J.~Yamagishi, and J.~S. Chung, ``To what extent can {ASV} systems naturally defend against spoofing attacks?'' in \emph{Interspeech 2024}, 2024, pp. 3240--3244.

\bibitem{jung2022sasv2022spoofingawarespeaker}
J.~weon Jung, H.~Tak, H.~jin Shim, H.-S. Heo, B.-J. Lee, S.-W. Chung, H.-J. Yu, N.~Evans, and T.~Kinnunen, ``{SASV} 2022: The first spoofing-aware speaker verification challenge,'' in \emph{Interspeech 2022}, 2022, pp. 2893--2897.

\bibitem{ggemini}
T.~Liu, K.~A. Lee, Q.~Wang, and H.~Li, ``Golden gemini is all you need: Finding the sweet spots for speaker verification,'' \emph{IEEE/ACM Transactions on Audio, Speech, and Language Processing}, vol.~32, pp. 2324--2337, 2024.

\bibitem{Desplanques-2020}
\BIBentryALTinterwordspacing
B.~Desplanques, J.~Thienpondt, and K.~Demuynck, ``{ECAPA-TDNN}: Emphasized channel attention, propagation and aggregation in tdnn based speaker verification,'' in \emph{Interspeech 2020}, ser. interspeech-2020.\hskip 1em plus 0.5em minus 0.4em\relax ISCA, Oct. 2020. [Online]. Available: \url{http://dx.doi.org/10.21437/Interspeech.2020-2650}
\BIBentrySTDinterwordspacing

\bibitem{jung2021aasistaudioantispoofingusing}
J.-w. Jung, H.-S. Heo, H.~Tak, H.-j. Shim, J.~S. Chung, B.-J. Lee, H.-J. Yu, and N.~Evans, ``{AASIST}: Audio anti-spoofing using integrated spectro-temporal graph attention networks,'' in \emph{ICASSP 2022 - 2022 IEEE International Conference on Acoustics, Speech and Signal Processing (ICASSP)}, 2022, pp. 6367--6371.

\bibitem{tak2021end}
H.~Tak, J.~weon Jung, J.~Patino, M.~Kamble, M.~Todisco, and N.~Evans, ``{End-to-End Spectro-Temporal Graph Attention Networks for Speaker Verification Anti-Spoofing and Speech Deepfake Detection},'' in \emph{Proceedings of Interspeech 2021}, September 2021.

\bibitem{borodin24_asvspoof}
K.~Borodin, V.~Kudryavtsev, D.~Korzh, A.~Efimenko, G.~Mkrtchian, M.~Gorodnichev, and O.~Y. Rogov, ``{AASIST3: KAN-enhanced AASIST speech deepfake detection using SSL features and additional regularization for the ASVspoof 2024 Challenge},'' in \emph{The Automatic Speaker Verification Spoofing Countermeasures Workshop (ASVspoof 2024)}, 2024, pp. 48--55.

\bibitem{ge2022potentialjointlyoptimised}
W.~Ge, H.~Tak, M.~Todisco, and N.~Evans, ``On the potential of jointly-optimised solutions to spoofing attack detection and automatic speaker verification,'' in \emph{IberSPEECH 2022}, 2022, pp. 51--55.

\bibitem{sahidullah2016integrated}
M.~Sahidullah, H.~Delgado, M.~Todisco, H.~Yu, T.~Kinnunen, N.~Evans, and Z.-H. Tan, ``{Integrated Spoofing Countermeasures and Automatic Speaker Verification: An Evaluation on ASVspoof 2015},'' in \emph{Proceedings of the 17th Annual Conference of the International Speech Communication Association (Interspeech)}, 2016, pp. 1700--1704.

\bibitem{avishai}
A.~Weizman, Y.~Ben-Shimol, and I.~Lapidot, ``{Spoofing-Robust Speaker Verification Based on Time-Domain Embedding},'' in \emph{Cyber Security, Cryptology, and Machine Learning}, S.~Dolev, M.~Elhadad, M.~Kuty{\l}owski, and G.~Persiano, Eds.\hskip 1em plus 0.5em minus 0.4em\relax Cham: Springer Nature Switzerland, 2025, pp. 64--78.

\bibitem{weizman2024tandemspoofingrobustautomaticspeaker}
\BIBentryALTinterwordspacing
------, ``Tandem spoofing-robust automatic speaker verification based on time-domain embeddings,'' 2024. [Online]. Available: \url{https://arxiv.org/abs/2412.17133}
\BIBentrySTDinterwordspacing

\bibitem{probfusi}
Y.~Zhang, G.~Zhu, and Z.~Duan, ``A probabilistic fusion framework for spoofing aware speaker verification,'' in \emph{The Speaker and Language Recognition Workshop (Odyssey 2022)}, 2022, pp. 77--84.

\bibitem{choi22b-interspeech}
J.-H. Choi, J.-Y. Yang, Y.-R. Jeoung, and J.-H. Chang, ``{HYU Submission for the SASV Challenge 2022: Reforming Speaker Embeddings with Spoofing-Aware Conditioning},'' in \emph{Interspeech 2022}, 2022, pp. 2873--2877.

\bibitem{Liu-2024}
\BIBentryALTinterwordspacing
X.~Liu, M.~Sahidullah, K.~A. Lee, and T.~Kinnunen, ``Generalizing speaker verification for spoof awareness in the embedding space,'' \emph{IEEE/ACM Transactions on Audio, Speech, and Language Processing}, vol.~32, p. 1261–1273, 2024. [Online]. Available: \url{http://dx.doi.org/10.1109/TASLP.2024.3358056}
\BIBentrySTDinterwordspacing

\bibitem{pact}
C.~Zhang and P.~C. Woodland, ``{DNN} speaker adaptation using parameterised sigmoid and {ReLU} hidden activation functions,'' in \emph{2016 IEEE International Conference on Acoustics, Speech and Signal Processing (ICASSP)}, 2016, pp. 5300--5304.

\bibitem{Bell-2021}
\BIBentryALTinterwordspacing
P.~Bell, J.~Fainberg, O.~Klejch, J.~Li, S.~Renals, and P.~Swietojanski, ``Adaptation algorithms for neural network-based speech recognition: An overview,'' \emph{IEEE Open Journal of Signal Processing}, vol.~2, p. 33–66, 2021. [Online]. Available: \url{http://dx.doi.org/10.1109/OJSP.2020.3045349}
\BIBentrySTDinterwordspacing

\bibitem{Li2020JointDecision}
J.~{Li}, Z.~{Wu}, J.~{Dang}, and H.~{Li}, ``{Joint Decision of Anti-Spoofing and Automatic Speaker Verification by Multi-Task Learning With Contrastive Loss},'' \emph{IEEE Access}, vol.~8, pp. 58\,534--58\,542, 2020.

\bibitem{wang2020asvspoof2019largescalepublic}
\BIBentryALTinterwordspacing
X.~Wang, J.~Yamagishi, M.~Todisco, H.~Delgado, A.~Nautsch, N.~Evans, M.~Sahidullah, V.~Vestman, T.~Kinnunen, K.~A. Lee, L.~Juvela, P.~Alku, Y.-H. Peng, H.-T. Hwang, Y.~Tsao, H.-M. Wang, S.~L. Maguer, M.~Becker, F.~Henderson, R.~Clark, Y.~Zhang, Q.~Wang, Y.~Jia, K.~Onuma, K.~Mushika, T.~Kaneda, Y.~Jiang, L.-J. Liu, Y.-C. Wu, W.-C. Huang, T.~Toda, K.~Tanaka, H.~Kameoka, I.~Steiner, D.~Matrouf, J.-F. Bonastre, A.~Govender, S.~Ronanki, J.-X. Zhang, and Z.-H. Ling, ``{ASV}spoof 2019: A large-scale public database of synthesized, converted and replayed speech,'' \emph{Computer Speech \& Language}, vol.~64, pp. 101--114, 2020. [Online]. Available: \url{https://www.sciencedirect.com/science/article/pii/S0885230820300474}
\BIBentrySTDinterwordspacing

\bibitem{adcf}
H.-J. Shim, J.-W. Jung, T.~Kinnunen, N.~Evans, J.-F. Bonastre, and I.~Lapidot, ``{a-DCF}: an architecture agnostic metric with application to spoofing-robust speaker verification,'' in \emph{Odyssey 2024}, 06 2024.

\bibitem{Confidence_Intervals}
\BIBentryALTinterwordspacing
L.~Ferrer and P.~Riera, ``Confidence intervals for evaluation in machine learning.'' [Online]. Available: \url{https://github.com/luferrer/ConfidenceIntervals}
\BIBentrySTDinterwordspacing

\end{thebibliography}

\end{document}